# Evaluating software defect prediction performance: an updated benchmarking study

Libo Li[1, *], Stefan Lessmann[2], and Bart Baesens[3]

Abstract— *Accurately predicting faulty software units helps practitioners target faulty units and prioritize their efforts to maintain software quality. Prior studies use machine-learning models to detect faulty software code. We revisit past studies and point out potential improvements. Our new study proposes a revised benchmarking configuration. The configuration considers many new dimensions, such as class distribution sampling, evaluation metrics, and testing procedures. The new study also includes new datasets and models. Our findings suggest that predictive accuracy is generally good. However, predictive power is heavily influenced by the evaluation metrics and testing procedure (frequentist or Bayesian approach). The classifier results depend on the software project. While it is difficult to choose the best classifier, researchers should consider different dimensions to overcome potential bias.*

[1]Kent Business School, University of Kent, United Kingdom, L.Li-411@kent.ac.uk

[2]The School of Business and Economics, Humboldt-University of Berlin, Unter-den-Linden 6, 10099 Berlin, Germany, stefan.lessmann@hu-berlin.de.

[3]The Department of Decision Sciences and Information Management, KU Leuven, Naamsestraat 69, 3000 Leuven Belgium, bart.baesens@kuleuven.be.

[*] Corresponding author

## 1  Introduction

Predicting defective code in the software development process is a key aspect of software analytics. Software testing firm Tricentis estimated the cost of software bugs at $1.1 trillion in 2016 [1-3]. Quality assurance resources are usually limited to maintaining the software. Predicting faulty units accurately allows developers and managers to prioritize their actions in the software development cycle and to address these faults. In defective software, a faulty unit might result from various factors that are hard to detect using human processes such as code review. Given the large size, number of lines of code, and complexity of a typical software project, a much more scalable approach is needed. Many researchers have proposed quantitative approaches to this research problem. Thanks to the proliferation of IT artifacts, researchers can use increasing amounts of data and conduct a wide range of experiments to pursue a better solution.

The number of available software projects not only provides researchers with a good collection of datasets, but also encourages many different research areas. For instance, data quality may influence predictive accuracy. The design of the model-building and selection process determines the model's ability to predict faulty units. The strength of the predictive results may not always be identical, since the evaluation metrics used will have different underlying theoretical foundations. Therefore, different models may score differently with various metrics. These variables are important parameters when designing experimental studies.

The literature shows that many benchmarking studies use machine learning classifiers to predict faulty units [4]. One of the most cited studies is that by Lessmann et al. [4]. However, with recent advances in machine learning, statistics science and other disciplines, it is time to think about improving benchmarking studies. More specifically, our study includes these new dimensions:

Software fault datasets might exhibit different quality in terms of the number of faulty units. When faulty classes are extremely rare in the dataset, this does not mean these errors should be ignored, but rather that it is essential to capture them. This is known as the class imbalance problem, since there are many more faultless units than faulty ones. We use sampling techniques to address the class imbalance problem. Namely, when the proportion of faulty units is very low, we use oversampling and undersampling techniques.

Recently, the literature has discussed the development of many new classifiers. We update our choice of classifiers in view of this. Obviously, it is impossible to include all of them in our study; we therefore prioritize the most commonly used methods.

Many benchmarking studies use NASA MDP datasets. This project contains datasets with well-known merits but also data quality issues [3]. In our new benchmarking study, we also include datasets collected from the GitHub (www.github.com) repository to gain better insights [5, 6].

Software defect predictions generated by classifiers should be assessed in terms of accuracy. While many studies use the Receiver operating characteristic (ROC) curve [7], or the Area under the ROC curve (also known as the AUC value), we include an alternative, namely the H measure, to address the potential limitations of the AUC. Our alternative measure corrects the mishandling of computing misclassification costs in AUC.

When observing the numerical differences in prediction results resulting from the use of different metrics on various datasets, statistical tests check whether these differences are statistically significant. We update the test procedure in the literature to examine critically whether a classifier i, does indeed substantially outperform another classifier j. Despite the classical Friedman test followed by post-hoc tests, we include a new Bayesian test procedure to gain insights against its Frequentist counterparts.

In view of these aspects, this study updates the work of Lessmann et al. [4], and of many other authors. In what follows, section 2 provides an overview of the literature. Section 3 explains the research setup for this study. Section 4 discusses the experimental results. The paper concludes in section 5. For ease of discussion, we use the term "software defect" and "faulty units" interchangeably, ignoring their minor differences.

## 2 LITERATURE OVERVIEW

### 2.1 Fault predictions

We formally define the dataset used as: $D = [\boldsymbol{x_i}, y_i]$. For observation $i \in [1, n]$, the boldface $\boldsymbol{x_i} = [x_1, x_2 \ldots x_p]$ represents the vector of independent variables (IV). The dependent variable is (DV) $y_i \in [-1; 1]$, where "1" stands for faulty and "-1" stands for non-faulty. The classifiers are built on the dataset and aim to classify the faulty units, yielding classifier results $\hat{y}_i = f(\boldsymbol{x_i})$. As the job of a classifier is to find out whether an observation is faulty [8-11], this is often referred to as the classification model, or binary classification model given its binary outcome. Different classifiers consider functional mappings $f(\cdot)$ in various forms using different underlying principles and hypotheses [6, 12, 13]. For instance, a regression model from the generalized linear regression family assumes that the relation between DV and its IVs can be expressed as a linear combination with some forms of transformation, often in a logarithmic scale. A tree-based classifier on the other hand, uses entropy theory to develop its model.

## 2.2 Classification techniques

Currently there are two main classes of classification techniques in software defect prediction: the statistical approach and the machine-learning approach. While the statistical approach uses traditional statistical models, such as regression models, the machine learning approach uses methods adapted from other research fields [14-17]. For instance, decision tree models and their ensemble counterpart random forests often appear in benchmarking studies as machine-learning candidates. Researchers may argue that the boundary between the two approaches is unclear, as all quantitative approaches are potentially rooted in statistics science [18, 19].

We consider six main classes of techniques, namely Bayesian approaches, tree-based approaches, support vector machine approaches, neural network approaches, boosting approaches, and others. These six main approaches cover a wide range of techniques. Some of them have been used in different contexts. Others, such as Bagged multilayer perceptron artificial neural networks and Ridge regression, have not been covered in prior studies, to the best of our knowledge [4]. However, these methods have shown promising results in other settings. Some other methods, such as the Bayesian and tree-based approaches, have appeared in prior benchmarking studies. They are also included in this new benchmarking study as the baseline.

We are aware of the growing body of literature on new classification methods [20-22]. Among many others, many studies have considered the deep learning paradigm [23, 24]. Deep learning models are more computationally extensive and difficult to train than many other classifiers. While deep learners obtain promising results on large datasets, the cost of building such models is often prohibitively high. The literature also shows that classical approaches such as support vector machines (SVM) might perform as well as deep learners [24]. Researchers should be aware of the trade-off between using complex and simple models. In particular, many software repositories are limited in size, and thus produce datasets that might not be large enough to justify the usage of deep learners.

It is necessary to fine-tune some models to achieve good results [25-27]. While some classifiers used in the literature do not require tuning, others do. An ensemble model, which is based on a bag of learners (e.g. tree learners), can be fine-tuned, since the number of learners can be tuned. Similarly, a support vector machine or a penalized regression model have weighting parameters in their objective functions for regularization. The same classifier algorithm may not perform equally well with different parameter configurations. Hence, a model selection procedure is needed to find the optimal results [4, 25].

When the number of observations with one class value dominates the other, e.g. $|y = 1| \gg |y = -1|$, we consider it as a class imbalance problem. Some classifiers may find it difficult to learn from an imbalanced dataset [28-32]. When one type of event (faulty unit) is rare, some classifiers fail to build a model capturing the underlying data distribution, while others just fail to converge. Oversampling and undersampling techniques can address this issue [33, 34]. These techniques start from the original dataset, then add more minority class observations with oversampling, and/or remove some majority class observations via undersampling. Both method operate only on the dataset without the need to modify the classification algorithm. The synthetic minority oversampling technique (SMOTE) is a widely considered solution and has successfully improved the accuracy of classifiers [35, 36]. The SMOTE technique randomly draws nearest neighbor instances of the minority instance and interpolates samples based on the original data and the random nearest neighbors. Prior research has extended the idea of interpolation by considering both minority and majority classes, together with other shortcomings of SMOTE [37]. Although the sampling technique does not impact classifiers at the algorithm level, it also influences classifier performance and must be considered when developing a benchmarking study[32]. We use advanced sampling techniques to improve dataset usability. Instead of relying on classifiers themselves to deal with class imbalance and to make inferences about the data, our benchmark study uses advanced sampling techniques to preprocess the data. In this way, we obtain a clear view of classifier performance with regard to software defect prediction, without being influenced by class distribution.

Preprocessing data will impact classifier performance [3]. Besides the class imbalance problem, researchers may experience data quality issues, which may jeopardize their results [38]. More specifically, as many software defect prediction studies use variables extracted from the original code for the classification model, researchers should investigate the quality of the extracted data. The data may contain duplicate records, inconsistencies, and so on [32]. Data preprocessing is necessary to address these issues.

## 2.3 Model evaluation

A classification model can provide predictive estimates to forecast faulty units in software projects. The classification model takes software metrics as inputs and produces a quantitative measure representing the likelihood of errors. In a benchmarking study, researchers may obtain different predictive estimates [38, 39]. It is vital to evaluate the predictive estimates and quantify predictive accuracies [40-42].

The Receiver Operating Characteristic (ROC) curve [7, 43, 44] is a commonly used measure for software defect prediction [4, 45-48]. The space of the ROC curve, often known as the AUC (area under the curve) measures the ability of a classifier to discriminate between faulty and non-faulty modules. The statistical interpretation of the AUC shows that the AUC is a probability that a classification model ranks a randomly chosen faulty observation higher than a randomly chosen non-faulty observation. A higher AUC value suggests that the corresponding classification model may predict better than those with lower AUC values. By definition, an AUC value ranges from [0.5, 1].

While many studies adopt the AUC measure, other work shows its potential flaws. For example, the AUC metric uses different misclassification costs for different classifiers [49]. The misclassification cost, in the context of software fault prediction, is associated with the fact that classification errors may differ in importance. Classifying "faulty" as "non-faulty" has a different cost from classifying "non-faulty" as "faulty." Research has shown that when using the AUC metric, the misclassification cost is related to the classifier [50]. This means that the AUC metric uses different rules to measure classifier performance and hence, should not be considered as a coherent measure [51]. We propose using a new alternative metric, the "H-measure" to address these issues. We introduce a dedicated weighting function to adjust the evaluation of the misclassification cost.

While a given metric measures predictive accuracy, a classification model might produce different results for different datasets. A classification model performs well on one dataset might perform less well on another dataset [52, 53]. Thus, we need to develop a rigorous testing procedure to find out whether certain methods outperform the others [54].

Formally, the statistical comparison of classifier models uses statistical tests such as a Friedman test to identify whether classifiers perform differently [4]. The Friedman test is favored over other parametrical tests such as ANOVA because it relaxes assumptions on normality and so on [4]. When the test result is significant, post-hoc tests will provide pairwise comparisons, to see whether one classifier outperforms the rest. For a given evaluation metric, the average ranks of each classifier over all datasets show the relative strength of their predictive performance. The post-hoc tests check whether the difference in average ranks are great enough to conclude that the performances are significantly different.

Formally, given $k$ different classifiers and $N$ datasets, we consider two classifiers $i$ and $j$ with average rank $R_i$ and $R_j$ to perform differently when $|R_i - R_j| \geq critical\ distance$. We define the critical distance as

$$critical\ distance = q_\alpha \sqrt{\frac{k(k+1)}{6N}} \quad (1)$$

$q_\alpha$ is a test statistic related to the number of classifiers [55].

Alternatively, researchers might use clustering techniques to separate classifiers into different groups based on their performance. One such technique is the Scott-Knott test [56]. Although many studies have attempted to use this method [33, 38, 41], the literature shows that the violation of normality assumption in this test and its potential extension will lead to statistical bias [57].

The approach using critical distance undoubtedly shows its popularity, and it has been used in many previous studies[4]. Nonetheless, this approach has still not addressed some important issues. For instance, in an experimental setup involving $k$ different classifiers and $N$ datasets, it may not make sense to include and compare all classifiers as some of them might underperform. More straightforwardly, when a classifier performs worse than other classifiers, it is less meaningful and may be removed from the comparison [45]. On the other hand, since the average ranks of two classifiers are tested based on the total number of classifiers $k$, having another "bad" classifier will influence the pairwise comparison, although it should not. Therefore, some researchers are in favor of pairwise comparisons rather than multiple comparison [58]. They argue that after an initial Friedman test, a Wilcoxon rank sum test can be used to compare two classifiers in a pairwise fashion, to help find the "best" classifier.

Another pitfall of such an approach is that most statistical

TABLE 1
AN ANALYSIS OF THE SOFTWARE DEFECT PREDICTION LITERATURE

| Title | year | Data | | | Models | | Evaluation | | Statistical tests | |
|---|---|---|---|---|---|---|---|---|---|---|
| | | Data quality | nr datasets | of Class Imbalance | classifiers | tuning | AUC | H | F test + post hoc | correction |
| [28] | 2018 | √ | 14 | √ | 7 | × | × | × | × | × |
| [33] | 2018 | √ | 9 | √ | 6 | √ | √ | × | × | × |
| [12] | 2017 | √ | 16 | × | 3 | / | √ | × | × | × |
| [54] | 2017 | × | 11 | × | 9 | × | √ | × | √ | × |
| [29] | 2017 | √ | 20 | √ | 5 | √ | × | × | × | × |
| [40] | 2016 | √ | 11 | × | 2 | × | √ | × | × | × |
| [30] | 2016 | √ | 16 | √ | 8 | √ | √ | × | × | × |
| [6] | 2016 | × | 15 | × | 13 | × | √ | × | × | × |
| [19] | 2016 | √ | 14 | × | 6 | √ | × | × | × | × |
| [41] | 2016 | √ | 18 | × | 3 | × | √ | × | × | √ |
| [25] | 2016 | √ | 17 | × | 4 | √ | √ | × | × | × |
| [26] | 2016 | √ | 18 | × | 26 | √ | √ | × | × | × |
| [27] | 2015 | × | 7 | × | 4 | √ | × | × | × | × |
| [31] | 2015 | √ | 7 | √ | 1 | √ | √ | × | × | × |
| [38] | 2015 | √ | 20 | × | 15 | ? | √ | × | × | √ |
| [32] | 2014 | √ | 10 | √ | 11 | × | √ | × | × | × |
| [14] | 2014 | ? | 9 | × | 5 | √ | √ | × | × | × |
| [52] | 2013 | × | 41 | × | 1 | √ | × | × | × | × |
| [45] | 2013 | × | 11 | × | 17 | √ | √ | √ | √ | × |
| [46] | 2013 | × | 9 | √ | 2 | × | √ | × | × | × |
| [34] | 2012 | × | 11 | √ | 2 | × | × | × | × | × |
| [47] | 2012 | √ | 3 | × | 4 | ? | √ | × | × | × |
| [15] | 2012 | ? | 10 | × | 4 | ? | √ | × | × | × |
| [53] | 2012 | √ | 34 | × | 5 | × | × | × | × | × |
| [20] | 2012 | × | 7 | × | 4 | ? | × | × | × | × |
| [48] | 2011 | √ | 17 | × | 3 | √ | √ | × | × | × |
| [21] | 2011 | × | 7 | × | 5 | ? | × | × | × | × |
| [42] | 2011 | ? | 3 | × | 1 | × | × | × | × | × |
| [22] | 2011 | × | 1 | × | 1 | / | × | × | × | × |
| [43] | 2010 | × | 10 | × | 9 | √ | √ | × | × | × |
| [8] | 2010 | √ | 3 | × | 1 | / | √ | × | × | × |
| [9] | 2010 | × | 1 | × | 7 | √ | √ | × | × | × |
| [39] | 2010 | × | 3 | × | 4 | × | × | × | × | × |
| [10] | 2009 | × | 10 | × | 1 | / | × | × | × | × |
| [13] | 2009 | × | 8 | × | 1 | / | √ | × | × | × |
| [11] | 2008 | × | 3 | √ | 7 | × | × | × | √ | × |
| [16] | 2008 | ? | 4 | × | 9 | × | × | × | × | × |
| [4] | 2008 | × | 10 | × | 22 | √ | √ | × | √ | × |
| [17] | 2007 | × | 6 | × | 3 | × | × | × | × | × |
| [44] | 2007 | × | 8 | × | 6 | × | √ | × | × | × |
| Average/count | | 17 | 11.3 | 9 | 6.2 | 15 | 24 | 1 | 4 | 2 |

√= yes, ×= no, = not mentioned or not clear, /= does not apply. More specifically, if a study has tuned the parameters, it is considered as "√", if the study did not tuned the parame-ters, or only those of some of the models, it is marked "×". "?" is used if the study does not mention the setting. "/" will be used if the standard does not apply, e.g. the model in use may not require tuning.
Unless otherwise noted, the number of datasets refers to different projects. Within the same project, there might be different dataset versions.
"AUC" refers to the area under the receiver operating characteristic curve. "H" refers to the H measure proposed in Hand's paper.
"F test" refers to the Friedman test, and "post-hoc" refers to post-hoc Nemenyi test.
"correction" refers to situations where the statistical tests have been treated to correct normality assumption, or other bias.

tests in use have been developed using a Frequentist approach. While much statistical science research has been moving towards the Bayesian paradigm, the software defect prediction literature has not yet entirely addressed this gap. Statistical tests such as ANOVA, the Friedman test and the Wilcoxon rank sum test are considered as Frequentist hypothesis testing procedures. Unfortunately, frequentist tests might not address the research needs. Researchers are interested in the power of the test, the probability of whether two classifier performances are identical or not, from the observed data. The confidence level does not link the probability to the observed data as such, but instead, provides a statement about the random draw of the sample data in general. A confidence level of 95% asserts that 19 out of 20 times, the estimated parameter might lie within the confidence interval of the collected sample, without knowing whether it lies within the particular observed data sample or not. A Bayesian approach, on the other hand, estimates the likelihood based on the observed data and calculates the posterior probability that one classifier will outperform the other. Furthermore, a Bayesian test allows us to compute the magnitude and uncertainty of the comparison. If one classifier is better than another, we know how strong the relationship is. Likewise, if we cannot conclude one is better than the other, we know to what extent they are identical, or namely "practically equivalent" [59]. While the Bayesian approach addresses some shortcomings of its Frequentist counterpart, it should be considered as an alternative rather than the panacea, since each paradigm has its own scope and limitations.

We summarize our literature analysis in Table 1. The inclusion rule takes into account the advancement of our prior study [60]. We have considered major software engineering outlets and conferences. Additionally, we have analyzed a number of review articles to broaden the horizon of the article search. A wide range of intellectual contribution should be taken into consideration, including but not limited to new classification methods, processing techniques, and tuning methods from various scholars. However, although between 70 and 100 articles might deserve to be included [61, 62], it is impossible to compare all of them for extensively. We will focus on the recent literature, since recent research benefits from greater data availability and methodological advancement than older work. For instance, the discussion of the problems of using Area under ROC curve only started around 2009 [51]. The same observation might be made for data collection from open source platforms such as GitHub. Moreover, journals often publish more content on this topic than conference publications, due to different word-count limits and expectations for research output. Thus, we summarize what we consider a representative list of publications.

On average, prior studies use 11.3 project datasets, and we observe that the number of datasets increases over the years, thanks to the proliferation of open source projects and metric extraction tools. Researchers identify and address data quality issues. Many studies address the class imbalance problem [4, 28, 29, 33]. Other studies do not, even though they mention the issue [12, 26]. The number of classifiers used is 6.2. This number is not considered large [4], because researchers tend to believe, in line with early results, that the choice of classifiers does not impact performance [38]. About 40% of the listed studies tune the model [25, 29] while others either do not [42, 53], or use default parameters [32, 39]. About 60% of the studies adopt the AUC metric, but very few of them address its potential limits [45]. To evaluate model performance, many studies use the t test [13, 26, 39] or the Scott-Knott test [38, 41], whose validity [4, 55, 57] has been questioned.

# 3 Research methods

## 3.1 Dataset collection

We have included the MDP project to evaluate classifier performance empirically (http://openscience.us/repo/defect/ ). The MDP project contains a number of defect datasets from many NASA artifacts, e.g. control software for observers, and has been used in many research articles[45]. While a popular choice for test datasets, discussions of data quality issues in the MDP project have attracted concern [3]. The literature reports a number of data quality issues that may jeopardize research outcomes, e.g. there are numerically identical variables, missing values, implausible values. To address this issue, we adopted the approach in [3] to improve data quality. While we retained the setup used in this prior study, we added one additional procedure to preprocess the dataset. Variables that are linear combinations of others may introduce a collinearity problem. While many software defect papers [4] do not discuss this issue specifically, the collinearity problem might introduce bias to the statistical model and influence the predictive outcome[2, 63]. We detected this problem using the "findLinearCombos" function in the R package "Caret"[64] and removed redundant linear combinations. We used a similar approach for datasets collected from the GitHub project.

While the MDP project is popular among researchers, public open source software repositories provide many more opportunities for empirical testing. We also included software defect datasets collected from GitHub [65]. This is an attractive data source, since commercial and/or confidential research projects often do not release their datasets. A version control system with bug-fixed reports allows the SZZ algorithm to automate the process of identifying software defects and constructing datasets [66]. It gives researchers more possibilities to collect datasets as they wish.

We captured the bugs in the GitHub dataset at both class and file level. We used class-level bug datasets. Investigating the file level could also be useful; however, a number of file-level datasets are limited in size and bug cases. For instance, all file-level datasets in the project "Android-Universal-Image-Loader" have fewer than 100 observations. Very small datasets cannot provide sufficient data for training and testing classifiers; hence, we chose class level datasets, as they are larger than their file-level counterparts are. In most software projects, one file contains exactly one class. However, multiple classes can appear in one file. The reason for this is that in some files, the class structures are nested, and inner classes may exist for various coding purposes and styles.

Unlike the MDP project, the GitHub project evolves over different versions, and thus has different waves of datasets. Since the number of different versions is large, we only present an aggregated average result here. The full result appears in the appendix.

We addressed the data imbalance problem by oversampling until the faulty class reached 20%, using Adaptive synthetic sampling (ADASYN) algorithm[37], because the mean and median defect rate of datasets that do not suffer from the imbalance problem is 17.8% and 18 % respectively; so roughly 20%. While the literature shows that a balanced class distribution may lead to good classification results, a minority class ratio of 20% also yields promising results [67]. Another reason is that empirically we do not expect the bug rate to be very high; so 20% would be a large enough number. It would be very unlikely for 50% of the code in a dataset to be faulty. An overview of the datasets appears in Table 2.

TABLE 1

AN OVERVIEW OF THE DATASETS

| MDP | Number of observations | Number of variables | Number of fault observations | Percentage of fault |
|---|---|---|---|---|
| CM1 | 688 | 37 | 84 | 12.21% |
| JM1 | 19186 | 21 | 3518 | 18.34% |
| KC1 | 4192 | 21 | 650 | 15.51% |
| KC3 | 400 | 39 | 72 | 18.00% |
| MC1 | 18554 | 38 | 136 | 0.73% |
| MC2 | 254 | 39 | 88 | 34.65% |
| MW1 | 528 | 37 | 54 | 10.23% |
| PC1 | 1518 | 37 | 122 | 8.04% |
| PC2 | 3170 | 36 | 32 | 1.01% |
| PC3 | 2250 | 37 | 280 | 12.44% |
| PC4 | 2798 | 37 | 356 | 12.72% |
| PC5 | 34002 | 38 | 1006 | 2.96% |
| **GitHub (aggregated average)** | | | | |
| Android-Universal-Image-Loader | 124.80 | 73.80 | 31.40 | 25.16% |
| BroadleafCommerce | 1714.60 | 91.20 | 137.20 | 8.00% |
| MapDB | 496.80 | 86.40 | 106.40 | 21.42% |
| antlr4 | 587.40 | 86.20 | 127.40 | 21.69% |
| ceylon-ide-eclipse | 1469.00 | 87.50 | 200.00 | 13.61% |
| elasticsearch | 4597.83 | 90.42 | 340.50 | 7.41% |
| hazelcast | 2964.00 | 89.88 | 174.38 | 5.88% |
| junit | 843.20 | 84.40 | 152.60 | 18.10% |
| mcMMO | 217.40 | 73.80 | 40.40 | 18.58% |
| mct | 2695.00 | 92.00 | 672.33 | 24.95% |
| neo4j | 5793.00 | 91.00 | 848.67 | 14.65% |
| netty | 1119.25 | 88.00 | 118.88 | 10.62% |
| orientdb | 1828.60 | 90.60 | 154.80 | 8.47% |
| oryx | 595.67 | 86.33 | 96.00 | 16.12% |
| titan | 1086.50 | 88.00 | 226.00 | 20.80% |

## 3.2 Benchmarking classifiers

We included 17 classifiers in this study. As discussed in the literature review, the selection includes six main classes of classifiers: Bayesian approaches, tree-based approaches, support vector machine approaches, neural network approaches, boosting approaches, and others. We adopted the Matlab, R and Weka implementation of those classifiers. Although classifiers with default parameter settings can predict defective units, we fine-tuned them to increase their predictive performance [25]. We considered the same classifier algorithm with a different parameter configuration as a different candidate model. For example, we considered CART models with different parameter values for "minleaf" (the minimal number of observations per tree leaf) as different candidate models. When testing a specific algorithm, we assessed candidate models with different parameters using cross validation within the training set. We used the candidate model with the best predictive performance for testing. We present an overview of all classifiers in TABLE .

TABLE 3
AN OVERVIEW OF THE CLASSIFIERS

| Classifier names | Acronym | Implementation | Candidate models |
|---|---|---|---:|
| Bagged multilayer perceptron artificial neural network | BaggingModelANN | Matlab | 4 |
| Boosted decision trees | BoostingModelAdaBoostM1 | Matlab | 9 |
| CART | CARTModel | Matlab | 12 |
| Logistic regression | LRModel | Matlab | 1 |
| Multilayer perceptron artificial neural network | MLPModel | Matlab | 171 |
| Random forest | RFModelR | R package "randomForest" | 35 |
| Ridge Regression | RidgeRegressionModel | Matlab | 10 |
| Linear support vector machine | SVMModelLibLinear | Matlab | 29 |
| SVM with radial basis kernel function | SVMModelRbf | Matlab | 300 |
| Alternating decision tree | WEKAModelADT | WEKA | 5 |
| Tree Augmented Naive Bayes | WEKAModelBayesNetTAN | WEKA | 1 |
| J4.8 | WEKAModelJ48 | WEKA | 12 |
| k-nearest neighbor | WEKAModelKnn | WEKA | 8 |
| Logistic model tree | WEKAModelLMT | WEKA | 1 |
| Naive Bayes | WEKAModelNaiveBayes | WEKA | 1 |
| Radial basis function neural network | WEKAModelRBFNetwork | WEKA | 5 |
| Voted perceptron | WEKAModelVP | WEKA | 1 |

3.3 Experimental setup

We split the datasets into training and testing sets using fivefold cross validation, to assess their predictive accuracy. Within each fold, we used another internal five-fold cross validation for model selection, to avoid the potential bias of training and testing models on the same dataset. We conducted this process to find the best parameter configuration for each classifier. We assessed predictive outcome using the AUC and the H measure. When calculating the H measure, we set the underlying Beta distribution parameters to the constant value two. The Beta distribution function served as a weighting function to address the shortcoming of the AUC metric, as is common in the literature [51, 60].

We evaluate the obtained results with average ranks first. For the top performing candidates, we did not only use post-hoc tests but also compared the post-hoc results using Bayesian tests. More specifically, we compared the top performing classifiers we compare them in a pairwise fashion manner. Classifier-performance comparisons often violate the assumption that the samples are independent and identically distributed (i.i.d). Bayesian tests are useful, since they build a hierarchical model based on the joint distribution learned from the sample. Furthermore, the posterior probabilities $P(classifier\ i \gg classifier\ j)$, $P(classifier\ i = classifier\ j)$ and $P(classifier\ i \ll classifier\ j)$ estimate whether one classifier outperforms another, or they are "practically equivalent," meaning that we cannot empirically conclude which is better. The "practically equivalent" situation happens when the mean difference of two classifiers lies in a very small region, such as [-0.01, 0.01] [59, 68]. This region is also known as a **r**egion **o**f **p**ractical **e**quivalence (rope) [69]. In other words, when the Bayesian hierarchical test results fall into the "rope", we consider the classifiers as practically equivalent. For our experimental setup, the $AUC_{rope} = [-0.01, 0.01]$ as suggested in the literature [59, 69]. However, there is a lack of prior evidence with regard to the H measure. After some experiments, we set $H_{rope} =$

$[-0.05, 0.05]$, as such a configuration is most stable when conducting Bayesian tests (https://github.com/BayesianTestsML). The reason for the difference between the AUC and H metrics is that the variance of possible AUC values is much smaller than the variance of the H measure, and thus should be adjusted differently for the "rope".

## 4 Results and discussion

In this section, we first report the results of the machine learning models and statistical comparison in section 4.1. In section 4.2, we report the findings of our study and discuss their differences from other work. We discuss limitations and future work in section 4.3.

### 4.1 Model results and statistical tests

We list the results of our experiments below. The GitHub project results are aggregated while the MDP project results are not. We have retained the raw result of the MDP project to compare with prior literature. The raw GitHub datasets generated too many observations and results to display in the paper. We have included raw results of the GitHub projects at different times in the Appendix. TABLE 2 and TABLE 3 report the AUC and H measure results of the MDP project, respectively.

TABLE 2
AUC RESULTS OF MDP DATASETS

|  | CM1 | JM1 | KC1 | KC3 | MC1 | MC2 | MW1 | PC1 | PC2 | PC3 | PC4 | PC5 | Average |
|---|---|---|---|---|---|---|---|---|---|---|---|---|---|
| BaggingModelANN | 0.984 | 0.737 | 0.863 | 0.955 | 0.992 | 0.948 | 0.966 | 0.981 | 0.994 | 0.955 | 0.986 | 0.981 | 0.945 |
| BoostingModelAdaBoostM1 | 0.926 | 0.733 | 0.840 | 0.955 | 0.987 | 0.961 | 0.926 | 0.952 | 0.984 | 0.904 | 0.970 | 0.977 | 0.926 |
| CARTModel | 0.381 | 0.507 | 0.500 | 0.312 | 0.403 | 0.233 | 0.407 | 0.397 | 0.360 | 0.385 | 0.484 | 0.500 | 0.406 |
| LRModel | 0.854 | 0.707 | 0.809 | 0.866 | 0.929 | 0.868 | 0.833 | 0.882 | 0.970 | 0.837 | 0.923 | 0.960 | 0.870 |
| MLPModel | 0.977 | 0.770 | 0.919 | 0.954 | 0.991 | 0.981 | 0.920 | 0.981 | 0.992 | 0.979 | 0.984 | 0.986 | 0.953 |
| RFModelR | 0.960 | 0.947 | 0.950 | 0.992 | 0.952 | 0.990 | 0.942 | 0.982 | 0.996 | 0.986 | 0.994 | 0.988 | 0.973 |
| RidgeRegressionModel | 0.830 | 0.709 | 0.808 | 0.842 | 0.916 | 0.785 | 0.816 | 0.873 | 0.902 | 0.837 | 0.906 | 0.954 | 0.848 |
| SVMModelLibLinear | 0.797 | 0.708 | 0.801 | 0.818 | 0.934 | 0.875 | 0.803 | 0.860 | 0.906 | 0.833 | 0.896 | 0.954 | 0.849 |
| SVMModelRbf | 0.966 | 0.843 | 0.903 | 0.994 | 0.982 | 0.993 | 0.904 | 0.949 | 0.992 | 0.989 | 0.984 | 0.984 | 0.957 |
| WEKAModelADT | 0.974 | 0.765 | 0.887 | 0.967 | 0.991 | 0.933 | 0.994 | 0.977 | 0.986 | 0.927 | 0.986 | 0.986 | 0.948 |
| WEKAModelBayesNetTAN | 0.814 | 0.727 | 0.821 | 0.797 | 0.973 | 0.785 | 0.807 | 0.884 | 0.943 | 0.839 | 0.922 | 0.977 | 0.857 |
| WEKAModelJ48 | 0.870 | 0.667 | 0.786 | 0.885 | 0.500 | 0.951 | 0.799 | 0.764 | 0.500 | 0.815 | 0.933 | 0.859 | 0.777 |
| WEKAModelKnn | 0.882 | 0.900 | 0.921 | 0.813 | 0.994 | 0.898 | 0.862 | 0.912 | 0.865 | 0.921 | 0.918 | 0.985 | 0.906 |
| WEKAModelLMT | 0.972 | 0.947 | 0.953 | 0.937 | 0.982 | 0.944 | 0.964 | 0.974 | 0.994 | 0.957 | 0.988 | 0.990 | 0.967 |
| WEKAModelNaiveBayes | 0.750 | 0.682 | 0.792 | 0.703 | 0.917 | 0.747 | 0.770 | 0.804 | 0.896 | 0.769 | 0.836 | 0.940 | 0.801 |
| WEKAModelRBFNetwork | 0.909 | 0.720 | 0.860 | 0.925 | 0.970 | 0.932 | 0.891 | 0.892 | 0.813 | 0.939 | 0.945 | 0.972 | 0.897 |
| WEKAModelVP | 0.749 | 0.664 | 0.726 | 0.780 | 0.652 | 0.789 | 0.772 | 0.775 | 0.552 | 0.762 | 0.835 | 0.817 | 0.739 |

TABLE 3
H MEASURE OF MDP DATASETS

|  | CM1 | JM1 | KC1 | KC3 | MC1 | MC2 | MW1 | PC1 | PC2 | PC3 | PC4 | PC5 | Average |
|---|---|---|---|---|---|---|---|---|---|---|---|---|---|
| BaggingModelANN | 0.802 | 0.136 | 0.322 | 0.821 | 0.504 | 0.777 | 0.701 | 0.698 | 0.727 | 0.603 | 0.750 | 0.389 | 0.603 |
| BoostingModelAdaBoostM1 | 0.778 | 0.124 | 0.265 | 0.805 | 0.521 | 0.876 | 0.749 | 0.586 | 0.939 | 0.343 | 0.643 | 0.348 | 0.581 |
| CARTModel | 0.000 | 0.007 | 0.000 | 0.001 | 0.000 | 0.006 | 0.000 | 0.000 | 0.000 | 0.000 | 0.000 | 0.000 | 0.001 |
| LRModel | 0.292 | 0.107 | 0.212 | 0.442 | 0.239 | 0.574 | 0.396 | 0.285 | 0.265 | 0.212 | 0.443 | 0.289 | 0.313 |
| MLPModel | 0.828 | 0.187 | 0.570 | 0.853 | 0.637 | 0.898 | 0.703 | 0.846 | 0.868 | 0.809 | 0.870 | 0.477 | 0.712 |
| RFModelR | 0.830 | 0.766 | 0.755 | 0.890 | 0.727 | 0.928 | 0.803 | 0.791 | 0.960 | 0.838 | 0.877 | 0.825 | 0.833 |
| RidgeRegressionModel | 0.244 | 0.107 | 0.217 | 0.365 | 0.224 | 0.401 | 0.321 | 0.235 | 0.124 | 0.217 | 0.393 | 0.272 | 0.260 |
| SVMModelLibLinear | 0.250 | 0.107 | 0.216 | 0.387 | 0.134 | 0.615 | 0.268 | 0.219 | 0.064 | 0.179 | 0.326 | 0.251 | 0.251 |
| SVMModelRbf | 0.721 | 0.436 | 0.653 | 0.905 | 0.736 | 0.920 | 0.720 | 0.794 | 0.960 | 0.847 | 0.824 | 0.810 | 0.777 |
| WEKAModelADT | 0.820 | 0.139 | 0.361 | 0.796 | 0.693 | 0.829 | 0.915 | 0.753 | 0.950 | 0.524 | 0.761 | 0.412 | 0.663 |
| WEKAModelBayesNetTAN | 0.166 | 0.121 | 0.216 | 0.347 | 0.280 | 0.431 | 0.388 | 0.311 | 0.150 | 0.173 | 0.373 | 0.372 | 0.277 |
| WEKAModelJ48 | 0.468 | 0.091 | 0.197 | 0.588 | 0.000 | 0.774 | 0.405 | 0.217 | 0.000 | 0.245 | 0.441 | 0.218 | 0.304 |
| WEKAModelKnn | 0.619 | 0.715 | 0.723 | 0.701 | 0.740 | 0.759 | 0.706 | 0.713 | 0.543 | 0.767 | 0.803 | 0.788 | 0.715 |
| WEKAModelLMT | 0.735 | 0.716 | 0.740 | 0.812 | 0.738 | 0.808 | 0.734 | 0.725 | 0.630 | 0.807 | 0.905 | 0.828 | 0.765 |
| WEKAModelNaiveBayes | 0.122 | 0.084 | 0.160 | 0.181 | 0.008 | 0.310 | 0.235 | 0.137 | 0.015 | 0.157 | 0.203 | 0.159 | 0.148 |
| WEKAModelRBFNetwork | 0.544 | 0.131 | 0.348 | 0.664 | 0.349 | 0.826 | 0.606 | 0.461 | 0.335 | 0.585 | 0.522 | 0.371 | 0.478 |
| WEKAModelVP | 0.136 | 0.070 | 0.165 | 0.282 | 0.103 | 0.387 | 0.217 | 0.180 | 0.000 | 0.117 | 0.352 | 0.195 | 0.184 |

We summarize AUC and H measure results for the GitHub project in TABLE 4 and TABLE 5.

In terms of numerical values, the CART model performs the worst, even after parameter tuning. This observation holds true for both the GitHub and MDP projects, and using either the AUC or H measure. Another tree-based learner, the J4.8 classifier, has similar results with low AUC and H measures for both projects. The logistic regression model performs relatively well under the AUC measure, but when evaluated using the H measure, its ranking drops. While the AUC and H measure are strongly correlated (correlation coefficient = 0.978 for MDP and 0.951 for GitHub), the H measure ranks the classifiers somewhat differently from the AUC. In TABLE 6, we summarize the average ranks of the classifiers.

The classifier Bagged neural network (BaggingModelANN) tends to score lower using the H measure than with the AUC. The opposite is true for the Logistic model tree (WEKAModelLMT) and Radial basis function neural network (WEKAModelRBFNetwork), as they score higher in terms of H measure. This outcome appears more often with the GitHub project than the MDP project.

We highlight the five top-performing classifiers in red. The random forest model (RFModelR) appears to be the best model in terms of its ranks. "BaggingModelANN" and "MLPModel" also perform quite well.

The Friedman tests performed over the GitHub and MDP project using the AUC and H measure show that the differences are significant (all four p values $\ll 0.001$).

TABLE 4
AUC RESULTS OF GITHUB DATASETS

|  | Android | Broadleaf | MapDB | antlr4 | ceylon | elasticsearch | hazelcast | junit | mcMMO | mct | neo4j | netty | orientdb | oryx | titan | Average |
|---|---|---|---|---|---|---|---|---|---|---|---|---|---|---|---|---|
| BaggingModelANN | 0.917 | 0.836 | 0.950 | 0.961 | 0.908 | 0.863 | 0.855 | 0.939 | 0.820 | 1.000 | 0.906 | 0.871 | 0.852 | 0.879 | 0.971 | 0.902 |
| BoostingModelAdaBoostM1 | 0.776 | 0.828 | 0.927 | 0.962 | 0.915 | 0.845 | 0.848 | 0.930 | 0.792 | 0.999 | 0.899 | 0.863 | 0.863 | 0.854 | 0.973 | 0.885 |
| CARTModel | 0.156 | 0.353 | 0.129 | 0.147 | 0.329 | 0.386 | 0.400 | 0.171 | 0.355 | 0.069 | 0.289 | 0.308 | 0.342 | 0.250 | 0.143 | 0.255 |
| LRModel | 0.890 | 0.725 | 0.904 | 0.946 | 0.802 | 0.806 | 0.767 | 0.904 | 0.755 | 0.993 | 0.822 | 0.806 | 0.767 | 0.825 | 0.946 | 0.844 |
| MLPModel | 0.902 | 0.792 | 0.933 | 0.955 | 0.895 | 0.832 | 0.831 | 0.931 | 0.802 | 1.000 | 0.844 | 0.852 | 0.837 | 0.841 | 0.969 | 0.881 |
| RFModelR | 0.937 | 0.829 | 0.937 | 0.969 | 0.919 | 0.868 | 0.856 | 0.940 | 0.842 | 0.999 | 0.873 | 0.873 | 0.862 | 0.878 | 0.972 | 0.904 |
| RidgeRegressionModel | 0.895 | 0.822 | 0.898 | 0.937 | 0.866 | 0.816 | 0.796 | 0.910 | 0.723 | 0.978 | 0.865 | 0.827 | 0.811 | 0.837 | 0.939 | 0.861 |
| SVMModelLibLinear | 0.875 | 0.799 | 0.899 | 0.894 | 0.875 | 0.768 | 0.793 | 0.858 | 0.780 | 0.942 | 0.826 | 0.788 | 0.811 | 0.850 | 0.889 | 0.843 |
| SVMModelRbf | 0.917 | 0.788 | 0.926 | 0.953 | 0.883 | 0.788 | 0.793 | 0.941 | 0.780 | 0.999 | 0.853 | 0.821 | 0.828 | 0.866 | 0.952 | 0.873 |
| WEKAModelADT | 0.939 | 0.825 | 0.955 | 0.968 | 0.892 | 0.847 | 0.838 | 0.934 | 0.799 | 0.999 | 0.889 | 0.874 | 0.859 | 0.874 | 0.972 | 0.898 |
| WEKAModelBayesNetTAN | 0.880 | 0.820 | 0.947 | 0.963 | 0.906 | 0.832 | 0.831 | 0.925 | 0.820 | 0.999 | 0.887 | 0.841 | 0.846 | 0.862 | 0.972 | 0.889 |
| WEKAModelJ48 | 0.907 | 0.660 | 0.905 | 0.952 | 0.776 | 0.638 | 0.647 | 0.829 | 0.722 | 0.992 | 0.746 | 0.751 | 0.719 | 0.798 | 0.950 | 0.799 |
| WEKAModelKnn | 0.887 | 0.794 | 0.895 | 0.939 | 0.894 | 0.818 | 0.800 | 0.930 | 0.769 | 0.996 | 0.853 | 0.818 | 0.827 | 0.835 | 0.953 | 0.867 |
| WEKAModelLMT | 0.911 | 0.779 | 0.927 | 0.948 | 0.856 | 0.829 | 0.820 | 0.904 | 0.790 | 0.999 | 0.856 | 0.858 | 0.828 | 0.837 | 0.967 | 0.874 |
| WEKAModelNaiveBayes | 0.866 | 0.773 | 0.885 | 0.929 | 0.861 | 0.776 | 0.776 | 0.875 | 0.756 | 0.934 | 0.832 | 0.728 | 0.802 | 0.804 | 0.892 | 0.833 |
| WEKAModelRBFNetwork | 0.879 | 0.761 | 0.923 | 0.965 | 0.893 | 0.774 | 0.780 | 0.929 | 0.726 | 0.997 | 0.774 | 0.795 | 0.788 | 0.842 | 0.953 | 0.852 |
| WEKAModelVP | 0.845 | 0.751 | 0.897 | 0.916 | 0.841 | 0.683 | 0.734 | 0.881 | 0.747 | 0.972 | 0.735 | 0.768 | 0.768 | 0.816 | 0.905 | 0.817 |

TABLE 5
H MEASURE OF GITHUB DATASETS

|  | Android | Broadleaf | MapDB | antlr4 | ceylon | elasticsearch | hazelcast | junit | mcMMO | mct | neo4j | netty | orientdb | oryx | titan | Average |
|---|---|---|---|---|---|---|---|---|---|---|---|---|---|---|---|---|
| BaggingModelANN | 0.696 | 0.258 | 0.749 | 0.791 | 0.532 | 0.286 | 0.254 | 0.660 | 0.380 | 0.990 | 0.564 | 0.400 | 0.304 | 0.584 | 0.827 | 0.552 |
| BoostingModelAdaBoostM1 | 0.494 | 0.256 | 0.712 | 0.822 | 0.558 | 0.220 | 0.201 | 0.682 | 0.386 | 0.996 | 0.535 | 0.385 | 0.252 | 0.504 | 0.827 | 0.522 |
| CARTModel | 0.006 | 0.000 | 0.003 | 0.005 | 0.021 | 0.001 | 0.000 | 0.002 | 0.008 | 0.003 | 0.001 | 0.001 | 0.000 | 0.004 | 0.006 | 0.004 |
| LRModel | 0.723 | 0.177 | 0.641 | 0.789 | 0.509 | 0.198 | 0.132 | 0.614 | 0.252 | 0.908 | 0.428 | 0.229 | 0.180 | 0.455 | 0.767 | 0.467 |
| MLPModel | 0.786 | 0.237 | 0.746 | 0.841 | 0.532 | 0.240 | 0.217 | 0.694 | 0.344 | 0.998 | 0.550 | 0.421 | 0.268 | 0.565 | 0.833 | 0.551 |
| RFModelR | 0.802 | 0.269 | 0.778 | 0.851 | 0.559 | 0.291 | 0.242 | 0.680 | 0.432 | 0.998 | 0.575 | 0.438 | 0.316 | 0.565 | 0.852 | 0.577 |
| RidgeRegressionModel | 0.670 | 0.240 | 0.678 | 0.746 | 0.442 | 0.196 | 0.172 | 0.577 | 0.308 | 0.803 | 0.395 | 0.293 | 0.235 | 0.531 | 0.688 | 0.465 |
| SVMModelLibLinear | 0.624 | 0.222 | 0.580 | 0.548 | 0.384 | 0.127 | 0.145 | 0.430 | 0.328 | 0.654 | 0.281 | 0.156 | 0.198 | 0.377 | 0.528 | 0.372 |
| SVMModelRbf | 0.755 | 0.238 | 0.723 | 0.820 | 0.545 | 0.213 | 0.171 | 0.700 | 0.351 | 0.980 | 0.518 | 0.330 | 0.248 | 0.570 | 0.766 | 0.524 |
| WEKAModelADT | 0.793 | 0.247 | 0.745 | 0.820 | 0.561 | 0.253 | 0.220 | 0.658 | 0.364 | 0.995 | 0.553 | 0.415 | 0.268 | 0.536 | 0.846 | 0.552 |
| WEKAModelBayesNetTAN | 0.704 | 0.251 | 0.778 | 0.839 | 0.579 | 0.195 | 0.195 | 0.644 | 0.425 | 0.997 | 0.548 | 0.303 | 0.245 | 0.499 | 0.836 | 0.536 |
| WEKAModelJ48 | 0.732 | 0.187 | 0.650 | 0.740 | 0.495 | 0.148 | 0.110 | 0.505 | 0.306 | 0.901 | 0.419 | 0.298 | 0.193 | 0.463 | 0.779 | 0.462 |
| WEKAModelKnn | 0.705 | 0.215 | 0.690 | 0.716 | 0.520 | 0.221 | 0.179 | 0.607 | 0.298 | 0.972 | 0.528 | 0.306 | 0.239 | 0.456 | 0.753 | 0.494 |
| WEKAModelLMT | 0.740 | 0.225 | 0.756 | 0.811 | 0.551 | 0.228 | 0.217 | 0.663 | 0.345 | 0.994 | 0.540 | 0.389 | 0.252 | 0.517 | 0.825 | 0.537 |
| WEKAModelNaiveBayes | 0.701 | 0.126 | 0.595 | 0.655 | 0.373 | 0.137 | 0.081 | 0.439 | 0.266 | 0.635 | 0.282 | 0.180 | 0.122 | 0.382 | 0.517 | 0.366 |
| WEKAModelRBFNetwork | 0.763 | 0.220 | 0.717 | 0.819 | 0.513 | 0.198 | 0.162 | 0.664 | 0.301 | 0.930 | 0.444 | 0.324 | 0.222 | 0.524 | 0.770 | 0.505 |
| WEKAModelVP | 0.589 | 0.190 | 0.635 | 0.657 | 0.428 | 0.148 | 0.112 | 0.497 | 0.269 | 0.760 | 0.379 | 0.195 | 0.152 | 0.390 | 0.595 | 0.400 |

We conducted the post-hoc analysis to examine whether there are individual differences between classifiers. The critical distance is calculated as indicated in equation 1. Recall the number of datasets in MDP and GitHub are 12 and 15 respectively. As mentioned earlier, we did not include all 17 classifiers in the post-hoc tests since classifiers such as CART performed poorly on all datasets. Including CART would "inflate" the classifier count $k$, the test statistics $q_\alpha$ and perhaps the critical distance. We give a numerical example below.

$$cd_{MDP} = 3.458\sqrt{\frac{17(17+1)}{6\times 12}} = 7.13$$

$$cd_{GitHub} = 3.458\sqrt{\frac{17(17+1)}{6\times 15}} = 6.38$$

When using all classifiers to compute the critical distance, the "biased" results suggest that only differences in ranks larger than 7.13 for MDP and 6.38 for the GitHub project should be considered significant.

TABLE 6
AVERAGE RANKS

|  | GitHub | | MDP | |
| --- | --- | --- | --- | --- |
|  | AUC | h | AUC | h |
| BaggingModelANN | 2.7 | 4.7 | 4.2 | 6.7 |
| BoostingModelAdaBoostM1 | 4.6 | 6.1 | 6.7 | 6.8 |
| CARTModel | 17.0 | 17.0 | 17.0 | 16.8 |
| LRModel | 12.3 | 11.9 | 10.8 | 11.3 |
| MLPModel | 6.1 | 4.1 | 4.2 | 4.0 |
| RFModelR | 2.4 | 1.7 | 2.8 | 1.8 |
| RidgeRegressionModel | 10.3 | 10.8 | 12.2 | 12.2 |
| SVMModelLibLinear | 12.0 | 14.3 | 12.4 | 12.7 |
| SVMModelRbf | 7.7 | 7.0 | 4.3 | 3.2 |
| WEKAModelADT | 3.1 | 4.1 | 4.5 | 5.0 |
| WEKAModelBayesNetTAN | 5.4 | 5.9 | 10.7 | 11.7 |
| WEKAModelJ48 | 13.7 | 11.8 | 12.8 | 12.5 |
| WEKAModelKnn | 9.5 | 10.0 | 7.8 | 5.6 |
| WEKAModelLMT | 8.0 | 5.9 | 3.8 | 3.8 |
| WEKAModelNaiveBayes | 13.5 | 15.1 | 14.5 | 15.5 |
| WEKAModelRBFNetwork | 10.4 | 8.7 | 8.9 | 8.3 |
| WEKAModelVP | 14.3 | 14.0 | 15.3 | 15.1 |

Under such test conditions, none of the classifiers can be clearly identified as the "best."

If we limit our scope to five classifiers, since we picked the five top-performing classifiers from each category in TABLE 6, the test results change:

$$cd_{MDP\prime} = 2.728\sqrt{\frac{5(5+1)}{6 \times 12}} = 1.76$$

$$cd_{GitHub\prime} = 2.728\sqrt{\frac{5(5+1)}{6 \times 15}} = 1.58$$

We can conclude using this approach that the random forest model outperforms the rest in terms of H measure for the GitHub project. For the rest of the comparison, although the random forest model ranks higher, we cannot statistically conclude that it outperforms the rest, because the difference in ranks is smaller than the critical distance.

While the above testing procedure has often been used in the literature [4], its power is often doubted, and a Bayesian alternative is proposed [59]. We include a full report of comparisons of the various classifiers in the Appendix. TABLE 7 summarizes the comparative results of the top-performing classifiers

TABLE 7
PAIRWISE COMPARISON USING THE BAYESIAN TESTS

| Classifier 1 | Classifier 2 | AUC_MDP | H_MDP | AUC_GitHub | H_GitHub |
|---|---|---|---|---|---|
| BaggingModelANN | BoostingModelAdaBoostM1 | pe | pe | pe | pe |
| BaggingModelANN | MLPModel | pe | MLPModel | pe | pe |
| BaggingModelANN | RFModelR | pe | RFModelR | pe | pe |
| BaggingModelANN | SVMModelRbf | pe | SVMModelRbf | pe | pe |
| BaggingModelANN | WEKAModelADT | pe | pe | pe | pe |
| BaggingModelANN | WEKAModelBayesNetTAN | BaggingModelANN | BaggingModelANN | pe | pe |
| BaggingModelANN | WEKAModelLMT | pe | WEKAModelLMT | pe | pe |
| BoostingModelAdaBoostM1 | MLPModel | MLPModel | MLPModel | pe | pe |
| BoostingModelAdaBoostM1 | RFModelR | RFModelR | RFModelR | pe | pe |
| BoostingModelAdaBoostM1 | SVMModelRbf | SVMModelRbf | SVMModelRbf | pe | pe |
| BoostingModelAdaBoostM1 | WEKAModelADT | pe | pe | pe | pe |
| BoostingModelAdaBoostM1 | WEKAModelBayesNetTAN | BoostingModelAdaBoostM1 | BoostingModelAdaBoostM1 | pe | pe |
| BoostingModelAdaBoostM1 | WEKAModelLMT | pe | WEKAModelLMT | pe | pe |
| MLPModel | RFModelR | pe | pe | pe | pe |
| MLPModel | SVMModelRbf | pe | pe | pe | pe |
| MLPModel | WEKAModelADT | pe | pe | pe | pe |
| MLPModel | WEKAModelBayesNetTAN | MLPModel | MLPModel | pe | pe |
| MLPModel | WEKAModelLMT | pe | pe | pe | pe |
| RFModelR | SVMModelRbf | pe | pe | pe | pe |
| RFModelR | WEKAModelADT | pe | RFModelR | pe | pe |
| RFModelR | WEKAModelBayesNetTAN | RFModelR | RFModelR | pe | pe |
| RFModelR | WEKAModelLMT | pe | pe | pe | pe |
| SVMModelRbf | WEKAModelADT | pe | SVMModelRbf | pe | pe |
| SVMModelRbf | WEKAModelBayesNetTAN | SVMModelRbf | SVMModelRbf | pe | pe |
| SVMModelRbf | WEKAModelLMT | pe | pe | pe | pe |
| WEKAModelADT | WEKAModelBayesNetTAN | WEKAModelADT | WEKAModelADT | pe | pe |
| WEKAModelADT | WEKAModelLMT | pe | pe | pe | pe |
| WEKAModelBayesNetTAN | WEKAModelLMT | WEKAModelLMT | WEKAModelLMT | pe | pe |

*pe = Practically equivalent*

According to TABLE 7 indicates that we cannot identify a "best" classifier when using results from the GitHub study. The top performers selected from TABLE 6 are evaluated as "Practically equivalent." When observing the test results of the MDP study, the random forest model no longer outperforms the MLP model in terms of H measure. This means that although the random forest model is ranked higher than the others, there is insufficient evidence that it outperforms the rest. In general, Bagged multilayer perceptron artificial neural network, Multilayer perceptron artificial neural network, Random forest, and Alternating decision tree all perform quite well in the experiments but in no specific order. While we have no clear view of the best classifier, statistical tests still indicate significant differences between classifiers. Classifiers such as CART, Logistic regression, Ridge Regression, linear SVM, Naïve Bayes, Radial basis function neural network, and Voted perceptron are not as effective as the Bagged multilayer perceptron artificial neural network , Multilayer perceptron artificial neural network , Random forest and Alternating decision tree classifiers, as indicated by the Bayesian test results.

4.2 Comparison with other studies

The research results update our findings in a prior study [4] as we identify a list of classifiers that are worse than others. Other studies have also reported that the classifiers can be divided into two groups with regard to their performance [38]. Such a "divide" in classifier performance can be observed in the MDP project and the GitHub projects. The benchmark study by Lessmann et al. has been cited over seven hundred times, and a quick search of the cited papers yields many interesting citations. Quite often researchers propose one specific use of a machine learning model in software defect prediction, such as Naïve Bayes [10], and use the work of Lessmann et al. [4] to justify their choice of classifier(s) [53]. Most frequently, authors reason that since Lessmann et al. did not find one best classifier, it does not matter which classifier we use. A few researchers perceive the work of Lessmann et al. differently. For instance, Bennin et al. state that "*Lessman et al. [38] showed that RF was significantly better than 21*

*other prediction models.*"[29], although Lessmann et al. do not make any such assertion [4]. This study shows that, although it is still unclear which classifier performs the best, researchers should justify the use and validity of their choice of classifier [48]. They should also justify the reasoning behind their experimental setup and reporting [70].

A prior study reported that different empirical model validation methods introduce different levels of estimation bias, and that Single repetition holdout performs poorly as a validation method [41]. In some studies, when the researchers change the test setting from single repetition holdout to cross validation, the AUC score of certain classifiers changes drastically, and hence their ranking also changes. Yu et al. reported that Naïve Bayes performs better than Logistic regression and KNN when using a 50% training set. The same study also reported that when using 10-fold cross validation, Logistic regression and KNN outperform Naïve Bayes [12]. One cause of such confounded results is that the number of candidate models is limited. More extensive testing may avoid such bias. In our study, we observe that the CART model, which predicted better than random

[4] (AUC > 0.5) in the previous study when using the holdout method, performs poorly when fivefold cross validation is used. The average ranks of the CART model and the Voted perceptron model ("WEKAModelVP") are consistent, thanks to the large pool of classifiers used.

The GitHub datasets complement those of the MDP project interestingly. The GitHub results differ from those of the MDP project: a classifier that performs well on the GitHub project does not necessarily perform well on the MDP project. The finding might be interesting to software developers if they find their work similar to the open source projects on GitHub. The GitHub datasets include many different types of projects, such as data engines, business information systems, language processing tools, and games [6]. On the other hand, linking a specific software development project to the MDP datasets may not be easy. This is one benefit of using open source platforms as a data source for our study. As the MDP datasets are still being used in much research work, researchers and practitioners should consider whether the research findings derived from MDP are generalizable to their own software.

Another benefit of using open source software is that we can understand data quality problems better, since we have access to the source code. Prior research observed the data quality issue in MDP projects [3], but the reason for the problem was unclear. When the datasets contain duplicated observations, researchers do not have sufficient information to conclude whether these two observations represent two code modules with identical values, or a replication error. This is not the case with open source code, since each data observation is linked to its class and/or file.

### 4.3 Limitations and future studies

In this paper, we primarily focus on the binary classification of software defects. However, software defects can be predicted in many other ways. Future studies should focus on other predictive tasks, such as time taken to fix a bug [71]. In a time series setting, alternative measures should be considered to enrich the findings.

The literature has reported that class imbalance handling methods, e.g. sampling methods, will increase predictive accuracy [29, 33]. While in our study we consider sampling to improve data usability, as indicated in 3.1, we do not tune the sampling technique extensively to increase AUC or H metric scores, for instance by testing the class ratio. Due to the large number of classifiers and datasets, it is difficult to complete these experiments within a reasonable time. Furthermore, while the aim of software defect prediction is to build predictive models that identify defective units, model comprehensibility is also important, as we might want to understand how software features relate to defects. Sampling techniques change the class distribution, and hence, reduce the model's comprehensibility.

We used the GitHub repository for our benchmarking study as an open source software platform. Although the recent acquisition of GitHub [72] does not impact our research findings, it does raise concerns about whether the management style and user profile will remain the same in the future. These factors will affect software quality and defect datasets, and perhaps predictive accuracy.

# 5 Conclusion

We conducted our study using 17 classifiers on 27 datasets. Our work extends the literature [4, 61, 73] and includes a number of new dimensions. Our benchmarking study shows that software defect prediction should be assessed using extensive evaluation metrics and statistical tests. We discover that the random forest model (RFModelR) and neural network model (MLP) achieved quite good results. However, neither AUC nor H measure values lead to a significant difference in classifier performance. Meanwhile, it is quite complex to train and fine-tune these models. Therefore, the benefit of using complex models rather than simpler approaches is unclear [74]. If the predictive accuracy of a complex model such as the random forest is similar to that of simple models such as those from the WEKA library, then it might be better to choose a simpler model for practical reasons. Additionally, we notice that the AUC and H measures report different classifier performances. While the AUC measure is widely used in many studies [4, 24, 25, 45], an alternative metric can provide additional insights.

Stakeholders in software development are interested in using data science to make better decisions about their code. It is vital for researchers and practitioners to understand that advances in data science could affect their decisions. Particularly in the case of software defect prediction, our study shows that benchmarking study results should consider multiple dimensions, including the nature of datasets, predictive models, and evaluation procedures. It is critical to take advantage of new research findings to continue to improve defect prediction results.

Benchmarking study results depend heavily on the choice of statistical procedures. In addition to the choice of classifiers, datasets and evaluation metrics, the statistical test procedure might also affect the research findings. The Frequentist and Bayesian paradigms analyze the data in different ways. Each paradigm has its own strengths and weaknesses. The Bayesian paradigm addresses many problems in the Frequentist approach. For example, it reports posterior probability on empirical data, and relaxes the assumptions of statistical tests. The Bayesian approach is computationally intensive when there is a large number of datasets and classifiers. The underlying Bayesian Hierarchical model might not be a perfect choice to model the mean difference of classifier performances. It often takes several iterations and research attempts to find the best underlying model structure in a Bayesian setting [75]. While the Bayesian approach criticizes the Frequentist approach with its i.i.d. assumptions, the Bayesian approach also makes a moderate number of assumptions about parameter distribution [59]. When designing the tests, one should consider these limitations and the tradeoff between them. The comparison results from the two paradigms should be examined critically.

Essentially, benchmarking software defect prediction is an evergreen research topic[65]. Researchers may propose many different approaches to investigate this subject. The choice of classifiers can be different as there can be many novel classifiers. The availability of open source data allows benchmarking models to be tested on multiple datasets. It also makes it difficult for software developers to address data quality issues.

# 6   Acknowledgement

This work was partially supported by the French HPC Center ROMEO and NEOMA Business School.